\documentclass[epj]{webofc}
\usepackage[utf8]{inputenc}
\usepackage[varg]{txfonts}   
\usepackage{booktabs}
\usepackage{xcolor}
\definecolor{darkred}{rgb}{0.4,0.0,0.0}
\definecolor{darkgreen}{rgb}{0.0,0.4,0.0}
\definecolor{darkblue}{rgb}{0.0,0.0,0.4}
\usepackage[bookmarks,linktocpage,colorlinks,
    linkcolor = darkred,
    urlcolor  = darkblue,
    citecolor = darkgreen]{hyperref}
\usepackage{subfigure}
\wocname{EPJ Web of Conferences}
\woctitle{Lattice2017}
\newcommand{\I}{\ensuremath{\mathrm{i}}}

\newcommand{\tr}{\ensuremath{\mathrm{tr}}}
\newcommand{\aetap}{\text{a--}\eta'}
\newcommand{\api}{\text{a--}\pi}

\usepackage{hyperref}
\newcommand{\arxiv}[2]{[arXiv:\,\href{http://arxiv.org/abs/#1}{\texttt{#1}} [\texttt{#2}]]}
\newcommand{\arxivold}[1]{[arXiv:\,\href{http://arxiv.org/abs/#1}{\texttt{#1}}\,]}
\begin{document}
%
\selectlanguage{english}
\title{%
Ward identities in $\mathcal{N}=1$ supersymmetric SU(3) Yang-Mills theory on the lattice
}
\author{%
\firstname{Sajid} \lastname{Ali}\inst{1}\fnsep\thanks{Speaker, E-mail: sajid.ali@uni-muenster.de} \and
\firstname{Georg} \lastname{Bergner}\inst{2} \and
\firstname{Henning}  \lastname{Gerber}\inst{1} \and
\firstname{Pietro}  \lastname{Giudice}\inst{1} \and
\firstname{Istvan}  \lastname{Montvay}\inst{3} \and
\firstname{Gernot}  \lastname{M\"unster}\inst{1} \and
\firstname{Stefano}  \lastname{Piemonte}\inst{4} \and
\firstname{Philipp}  \lastname{Scior}\inst{1}
}
\institute{%
Institut f\"ur Theoretische Physik, Universit\"at M\"unster, Wilhelm-Klemm-Str.~9, D-48149 M\"unster
\and
Theoretisch-Physikalisches Institut, Universit\"at Jena, Max-Wien-Platz 1, D-07743 Jena
\and
Deutsches Elektronen-Synchrotron DESY, Notkestr.~85, D-22603 Hamburg
\and
Institute for Theoretical Physics, Universit\"at Regensburg, Universit\"atsstr.~31, D-93053 Regensburg
}
\abstract{%
The introduction of a space-time lattice as a regulator of field theories breaks symmetries 
associated with continuous space-time, i.e.\ Poincar{\'e} invariance and supersymmetry. A non-zero 
gluino mass in the supersymmetric Yang-Mills theory causes an additional soft breaking of supersymmetry. 
We employ the lattice form of SUSY Ward identities, imposing that their continuum form would be 
recovered when removing the lattice regulator, to obtain the critical hopping parameter where broken 
symmetries can be recovered.
}
\maketitle
\section{Introduction}
Supersymmetry (SUSY) is an extended symmetry between fermions and bosons which is expected to emerge at very high energies. Among all possible motivations for such a prediction, the fact that fermions and bosons are naturally paired in supersymmetric theories provides an elegant solution to the problem of the cosmological constant without the need of fine-tuning. The relevance of SUSY is not limited to search of new physics beyond the Standard Model, but it is also connected to the understanding of the non-perturbative properties of quantum field theories.

Our project is focused on numerical simulations of $\mathcal{N}=1$ supersymmetric Yang-Mills theory regularized on the lattice. The model describes the strong interactions between gluons and gluinos, two quantum fields related to each other by a SUSY transformation. In our simulations, we employ the plaquette gauge action and clover improved Wilson fermions. In this contribution we analyze the ensembles generated with the gauge group SU(3), while previous investigations have been done with the gauge group SU(2), see for instance Ref.~\cite{Bergner2016}.

On the lattice, Lorentz invariance is explicitly broken, and therefore supersymmetry can be recovered only in the continuum limit. In addition, Wilson fermions solve the doubling problem by breaking chiral symmetry explicitly. While sixteen fermion species would definitively break the pairing between bosons and fermions required by SUSY, on the other hand the absence of chiral symmetry implies that the fermion mass renormalizes additively. In other words, zero bare fermion mass does not correspond to zero renormalized fermion mass. The fine-tuning of the gluino mass requires special care, since a non-vanishing gluino mass is also a source of soft supersymmetry breaking. As proven in Ref.~\cite{Curci:1986sm}, in the continuum limit both the supersymmetric and the chiral Ward identities are consistent, and all the continuum symmetries of $\mathcal{N}=1$ SYM are recovered simultaneously. In this contribution we present a study of the SUSY Ward identities on the lattice, and we show that our results support the existence of a unique massless critical point up to lattice discretization errors.

\section{$\mathcal{N}=1$ supersymmetric Yang-Mills theory}\label{intro}

The Lagrangian of $\mathcal{N}=1$ supersymmetric Yang-Mills theory (SYM) in the
continuum is given by

\begin{equation}
\mathcal{L}=-\frac 14 F^a_{\mu\nu} F^a_{\mu\nu} + \frac{\I}{2} \bar{\lambda}^a \gamma_\mu (\mathcal{D}_{\mu} \lambda)^a - \frac{m_g}{2} \bar{\lambda}^a \lambda^a,
\end{equation}
where 
\begin{equation}
F^a_{\mu\nu} = \partial_\mu A^a_\nu -\partial_\nu A^a_\mu - \I\, g_0\,f^a_{bc}[A^b_\mu,A^c_\nu],
\end{equation} 
is the field strength tensor built from the gauge field $A_{\mu}(x)$, which 
represents the ``gluon''. The gluino field $\lambda(x)$, the fermion superpartner of 
the gluon, is minimally coupled to the gauge field. The gluino satisfies the 
Majorana condition
\begin{equation}
\overline{\lambda}(x) = \lambda^T(x) \, C\,,
\end{equation}
where $C$ is the charge conjugation matrix, and transforms under the adjoint representation of the gauge group. The covariant derivative thus reads 
\begin{equation}
(\mathcal{D}_{\mu} \lambda)^a = \partial_{\mu} \lambda^{a} + g_0\,f^a_{bc} A^b_\mu \lambda^{c}\,.
\end{equation} 
A non-zero mass of the
gluino in supersymmetric Yang-Mills theory causes a soft breaking of supersymmetry. 
For vanishing gluino mass it is expected that supersymmetry is unbroken in the continuum~\cite{Witten:1982df}.

\subsection{SUSY Yang-Mills theory on the lattice}

The introduction of a space-time lattice regulator necessarily breaks supersymmetry
\cite{Bergner:2009vg}. Models that preserve part of an extended superalgebra are discussed 
in Ref.~\cite{Catterall:2009it}. In our simulations, we use the ``Curci-Veneziano'' lattice 
action \cite{Curci:1986sm} 
\begin{equation}
\label{action}
S = S_g + S_f\, ,
\end{equation}   
where the gauge part ($S_g$) of the full action is given by the standard Wilson action:
\begin{equation}
S_g = \frac{\beta}{2} \sum_x\sum_{\mu\neq\nu}
\left( 1 - \frac{1}{N_c} {\rm Re}\,\tr\, U_{\mu\nu}(x) \right).
\end{equation}
Here $\beta \equiv 2N_c/g^2$ is the bare lattice gauge coupling for the SU($N_c$) gauge
field and $U_{\mu\nu}(x)$ is the standard plaquette variable.
The fermion part of the action in equation (\ref{action}) is 
\begin{equation}
\begin{array}{rcl}
S_f &\equiv& \frac 12 \sum_x \left\{ \overline{\lambda}^a(x)\lambda^a(x)
-\kappa \sum_{\mu=1}^4 \left[\overline{\lambda}^a(x+\hat{\mu}) V_\mu^{ab}(x)(1+\gamma_\mu)\lambda^b(x)
+\overline{\lambda}^a(x) V_\mu^{abT}(x) (1-\gamma_\mu)\lambda^b(x+\hat{\mu}) \right] \right.\\[2mm]
& & + \left.\frac{\I}{4}g_0 \kappa c_{SW} \overline{\lambda}^a(x) \sigma_{\mu\nu} P^{(cl)}_{\mu\nu;ab}(x)\lambda^b(x) \right\}.
\end{array}
\end{equation}
Here $\kappa$ is called the hopping parameter, which is related to the bare gluino mass by $\kappa=1/(2m_g+8)$,
$\gamma_\mu$ denotes a Dirac matrix, and $\sigma_{\mu\nu}=-\frac{1}{2}[\gamma_\mu,\gamma_\nu]$. The links $V_\mu^{ab}(x)$ are the gauge field variables in the adjoint representation, obtained from links $U_\mu(x)$ in the fundamental representation by
\begin{equation}
V_\mu^{ab}(x) \equiv 2 \,\tr \left( U^\dagger_{\mu}(x) T^a  U_{\mu}(x) T^b \right) ,
\end{equation}
where $T^a$ are the group generators of SU($N_c$). Finally, the last term in the fermion action contains the clover-symmetrized lattice field strength tensor:
\begin{equation}
P^{(cl)}_{\mu\nu}(x) = \frac{1}{4a} \sum_{i=1}^{4} \frac{1}{2\I g_0a} 
\bigg( U^{(i)}_{\mu\nu}(x) - U^{(i)\dagger}_{\mu\nu}(x)\bigg)\,.
\end{equation}
The coefficient $c_{SW}$ is tuned up to one-loop order in perturbation theory to improve the convergence of on-shell observables to the continuum limit \cite{MUS}. 


\section{Supersymmetric Ward identities}

In the continuum, Ward identities are of the form
\begin{equation}\label{wi}	
\big\langle \big( \partial_\mu j^\mu(x) \big) Q(y) \big\rangle = - \bigg\langle  
\frac{\delta Q(y)}{\delta \bar{\varepsilon}(x)}\bigg\rangle.
\end{equation}
Here $j^\mu(x)$ is the Noether current,
$Q(y)$ is an insertion operator and $\bar{\varepsilon}(x)$ is a parameter of infinitesimal symmetry
or supersymmetry transformations.
The right hand side of equation (\ref{wi}) is a contact term, which is zero if 
$Q(y)$ is localized at space-time points different from $x$.

\begin{figure}[thb]
	\centering
	\includegraphics[width=0.8\textwidth,clip]{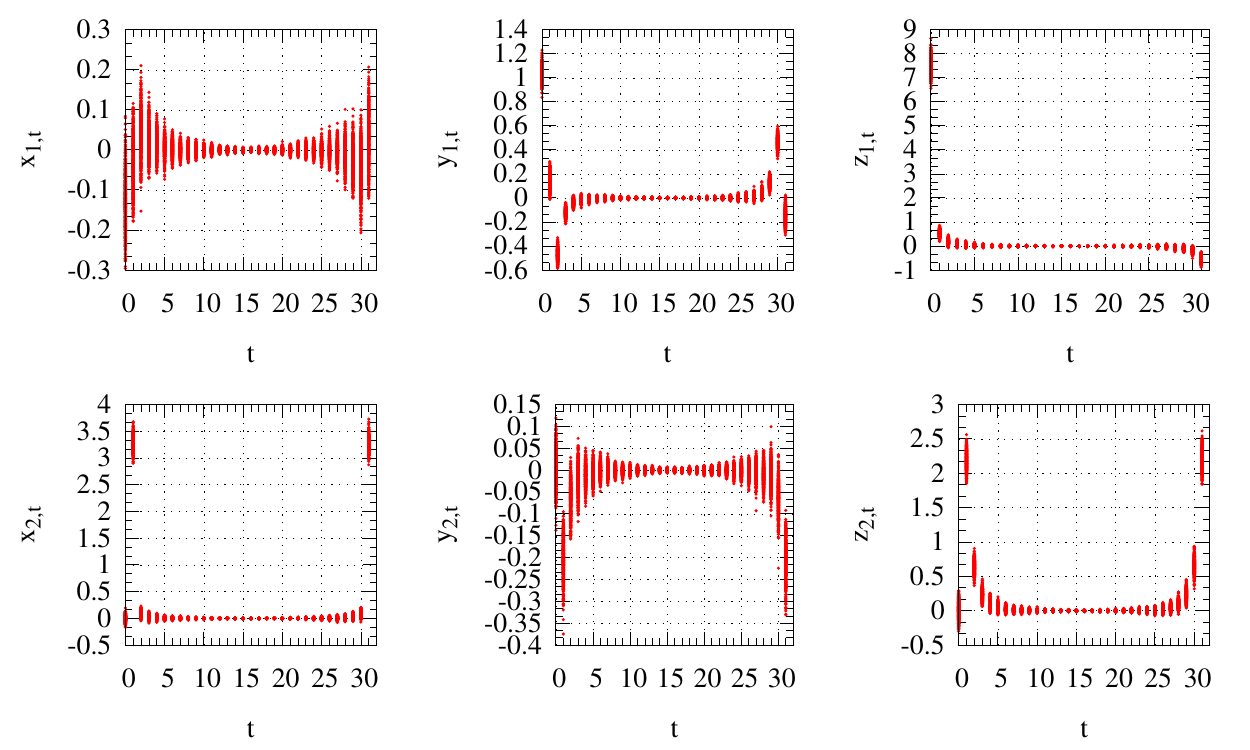}
	\caption{Numerical results for correlation functions at each time slice for gauge group SU(3), lattice volume, $V = 16^3 \cdot 32$, $\beta = 5.5$ and $\kappa = 0.1673$.}
	\label{corr}
\end{figure}

SUSY transformations on the lattice complying with parity ($\mathcal{P}$), charge conjugation ($\mathcal{C}$), 
time-reversal ($\mathcal{T}$) and gauge invariance can be defined by~\cite{Farchioni:2001wx}:
\begin{equation}
\begin{split}
\delta U_\mu(x) &= -\frac{\I g_0a}{2} \bigg( \bar{\epsilon}(x)\gamma_\mu U_\mu(x)\lambda(x) + 
\bar{\epsilon}(x+\hat{\mu})\gamma_\mu \lambda(x+\hat{\mu}) U_\mu(x)\bigg),\\
\delta U^\dagger_\mu(x) &= +\frac{\I g_0a}{2} \bigg( \bar{\epsilon}(x)\gamma_\mu \lambda(x) 
U^\dagger_\mu(x) + \bar{\epsilon}(x+\hat{\mu})\gamma_\mu U^\dagger_\mu(x) \lambda(x+\hat{\mu})\bigg),\\
\delta\lambda(x) &= + \frac{1}{2} P^{(cl)}_{\mu\nu}(x) \sigma_{\mu\nu} \epsilon(x),\\
\delta \bar{\lambda}(x) &= - \frac{1}{2}\bar{\epsilon}(x)\sigma_{\mu\nu}P^{(cl)}_{\mu\nu}(x).
\end{split}
\end{equation}
For any gauge invariant operator $Q(y)$ 
the above transformation results in the following Ward identities
\begin{equation}\label{wi2}
\sum_{\mu} \big\langle \big( \nabla_\mu S^{(sp)}_\mu(x) \big) Q(y)\big\rangle = 
m_0 \big\langle \chi(x)Q(y) \big\rangle + \big\langle X^{ps}(x) Q(y)\big\rangle - 
\bigg\langle  \frac{\delta Q(y)}{\delta \bar{\epsilon}(x)}\bigg\rangle,
\end{equation}
where
\begin{equation}
S^{(sp)}_\mu(x) = - \frac{1}{2} \sum_{\rho\sigma} \sigma_{\rho\sigma} \gamma_\mu 
Tr \bigg\{ P^{(cl)}_{\rho\sigma}(x) U^\dagger_\mu(x)\lambda(x+\hat{\mu}) U_\mu(x) + 
P^{(cl)}_{\rho\sigma}(x+\mu) U_\mu(x)\lambda(x) U^\dagger_\mu(x) \bigg\},
\end{equation}
\begin{equation}
\chi(x) = \sum_{\rho\sigma} \sigma_{\rho\sigma} Tr \bigg\{ P^{(cl)}_{\rho\sigma}(x) 
\lambda(x) \bigg\}.
\end{equation}
$S^{(sp)}_\mu(x)$ is the supercurrent.
In equation (\ref{wi2}) terms containing $\chi(x)$ and $X^{(ps)}(x)$ break SUSY. 
The first term $\chi(x)$ arises due to a non-zero bare gluino mass in the Lagrangian, and the term 
$X^{(ps)}(x)$ is introduced by the lattice regularization. At tree level $X^{(ps)}(x)$ is 
proportional to the lattice spacing $a$ and will vanish in the continuum limit. However, at 
higher orders $X^{(ps)}(x)$ has a finite contribution and restoration of supersymmetry is 
highly non-trivial. The renormalization of the supercurrent and of the gluino mass is therefore required. 
After renormalization the Ward identities get the following form:
\begin{equation}\label{wi3}
Z_S \big\langle \big( \nabla_\mu S_\mu(x) \big) Q(y)\big\rangle + Z_T \big\langle \big( 
\nabla_\mu T_\mu(x) \big) Q(y)\big\rangle = m_S \big\langle \chi(x)Q(y) \big\rangle + O(a) ,
\end{equation}
where $m_S = m_0 - a^{-1}Z_\chi$ is the subtracted mass, $Z_S$ , $Z_T$ and $Z_\chi$ are 
renormalization coefficients, and $T_\mu(x)$ is a mixing current.
Numerically it is convenient to use integrated Ward identities, where integration or summation 
is performed over three spatial coordinates. As a result of this integration, the Ward 
identity will hold on every time slice. Each term in the equation (\ref{wi3}) is a 4 $\times$ 4 matrix 
in Dirac space and can be expanded in the basis of $16$ Dirac matrices. Using discrete 
symmetries one can show that the surviving contributions form a set of two non-trivial independent equations~\cite{Farchioni:2001wx}:
\begin{equation}
\begin{split}
x_{1,t} + (Z_TZ^{-1}_S) y_{1,t} &= (am_SZ^{-1}_S) z_{1,t}, \\
x_{2,t} + (Z_TZ^{-1}_S) y_{2,t} &= (am_SZ^{-1}_S) z_{2,t}.
\end{split}
\end{equation}

\begin{equation}\label{setofequ}
\implies x_{b,t} + A y_{b,t} = B z_{b,t}, \quad b=1,2
\end{equation}
where
\begin{equation}
\begin{split}
x_{1,t} &\equiv \sum_{\vec{x}} \big\langle \nabla_4 S_4(x)\mathcal{O}(y) \big\rangle, \quad
x_{2,t} \equiv \sum_{\vec{x}}\big\langle \nabla_4 S_4(x) \gamma_4 \mathcal{O}(y) \big\rangle,\\
y_{1,t} &\equiv \sum_{\vec{x}} \big\langle \nabla_4 T_4(x) \mathcal{O}(y) \big\rangle, \quad 
y_{2,t} \equiv \sum_{\vec{x}} \big\langle \nabla_4 T_4(x)\gamma_4 \mathcal{O}(y) \big\rangle,\\
z_{1,t} &\equiv \sum_{\vec{x}} \big\langle \nabla_4 \chi(x) \mathcal{O}(y) \big\rangle, \quad 
z_{2,t} \equiv \sum_{\vec{x}} \big\langle \nabla_4 \chi(x)\gamma_4 \mathcal{O}(y) \big\rangle.
\end{split}
\end{equation}
These six different correlators are calculated numerically in Monte
Carlo simulations on Supercomputers, see \cite{Ali:2016zke}.
Figure \ref{corr} shows these correlators for one of our parameter sets.

\subsection{Global method}

To obtain estimates for $ A = Z_TZ^{-1}_S $ and $ B = am_SZ^{-1}_S $ from equation~(\ref{setofequ}), we minimize the quantity
\begin{equation}
\sum^{2}_{b=1} \sum^{t_{max}}_{t=t_{min}} \frac{(x_{b,t} + A y_{b,t} - B z_{b,t})^2}{\sigma_{b,t}^2}.
\end{equation} 
with respect to $A$ and $B$, where $\sigma_{b,t}^2$ is the 
sum of variances of these six correlators and can be calculated by using the Jackknife procedure.

\begin{figure}[thb]
    \centering
    \begin{minipage}{0.45\textwidth}
        \centering
        \includegraphics[width=1.1\linewidth,clip]{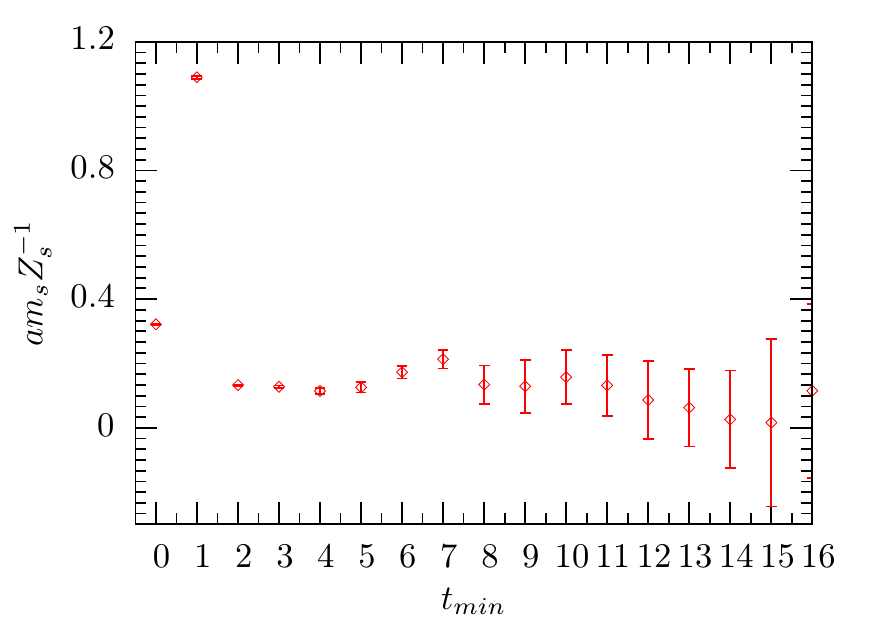}
        \caption{Gluino masses at each $t_{min}$ from the Global method.}\label{mg_gl}
    \end{minipage}\hspace{5mm}
    \begin{minipage}{0.45\textwidth}
        \centering
        \includegraphics[width=1.1\linewidth,clip]{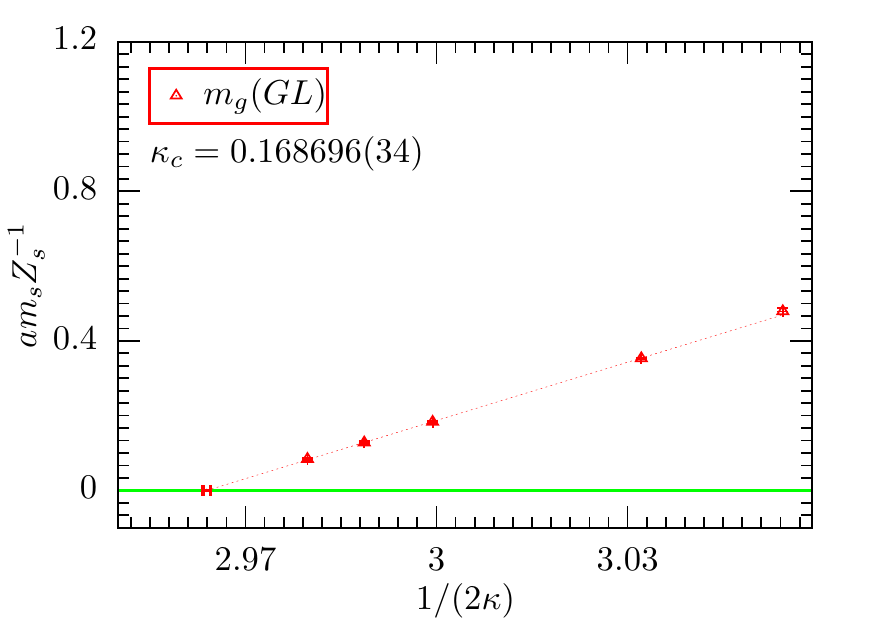}
        \caption{Extrapolation to the chiral limit using gluino masses from the Global method.}\label{Kc_gl}
    \end{minipage}
\end{figure}

Figure \ref{mg_gl} shows the results for $ B = am_SZ^{-1}_S $ from the Global method varying $t_{min}$. One 
can notice that the points which correspond to $t_{min}=0,1$ are off, which is due to contact terms. We take this into account by
considering only data at $t\geqslant3$. We developed two independent analysis codes in C/C++ and Octave, 
in order to be sure that the final results are correct.
We choose to take data of gluino masses at $t_{min}=3$ and repeat the procedure for the full range of $\kappa$, i.\,e.\ {0.1637, 0.1649, 0.1667, 0.1673 and 0.1678}, 
to get the extrapolation to $\kappa_c$ (figure \ref{Kc_gl}).
 

\subsection{Generalized least squares method}

The Global method, which has been discussed in the previous subsection, aims to find solutions for $A$ and $B$ 
numerically such that with the measured values $x_{b,t}, y_{b,t}, z_{b,t}$ the equations are satisfied 
approximately in an optimal way. This method, however, does not take into account the statistical correlations 
among the six operators at different time slices. We developed a method to improve on this point, so that
more reliable results and error estimates can be obtained.

Let's consider the equation (\ref{setofequ}) again which can also be written as
\begin{equation}
\sum_{\alpha} A_\alpha x_{i\alpha} = 0 ,
\end{equation}
where $A_\alpha$ for $\alpha=1,2,3$, are $1, A, -B$, whereas $x_{i\alpha}$ are $x_{b,t}, y_{b,t}, z_{b,t},$ with $i=(b,t)$. 
Employing the method of maximum likelihood \cite{Marshall:2005} we obtain the expression
\begin{equation}
L = \frac 12\sum_{i,\alpha,j,\beta} (A_\alpha \bar{x}_{i\alpha}) (D^{-1})_{ij} (A_\beta \bar{x}_{j\beta}).
\end{equation}
to be minimized, where
\begin{equation}
D_{ij} = \sum_{\alpha, \beta} A_\alpha A_\beta( \overline{x_{i\alpha} x_{j\beta}} - \bar{x}_{i\alpha} \bar{x}_{j\beta}).
\end{equation}
We have to find $A_\alpha$ numerically such that $L$ assumes its minimum. The details of the method will be explained in a follow-up paper.

\begin{figure}[thb]
    \centering
    \begin{minipage}{0.45\textwidth}
        \centering
        \includegraphics[width=1.15\linewidth,clip]{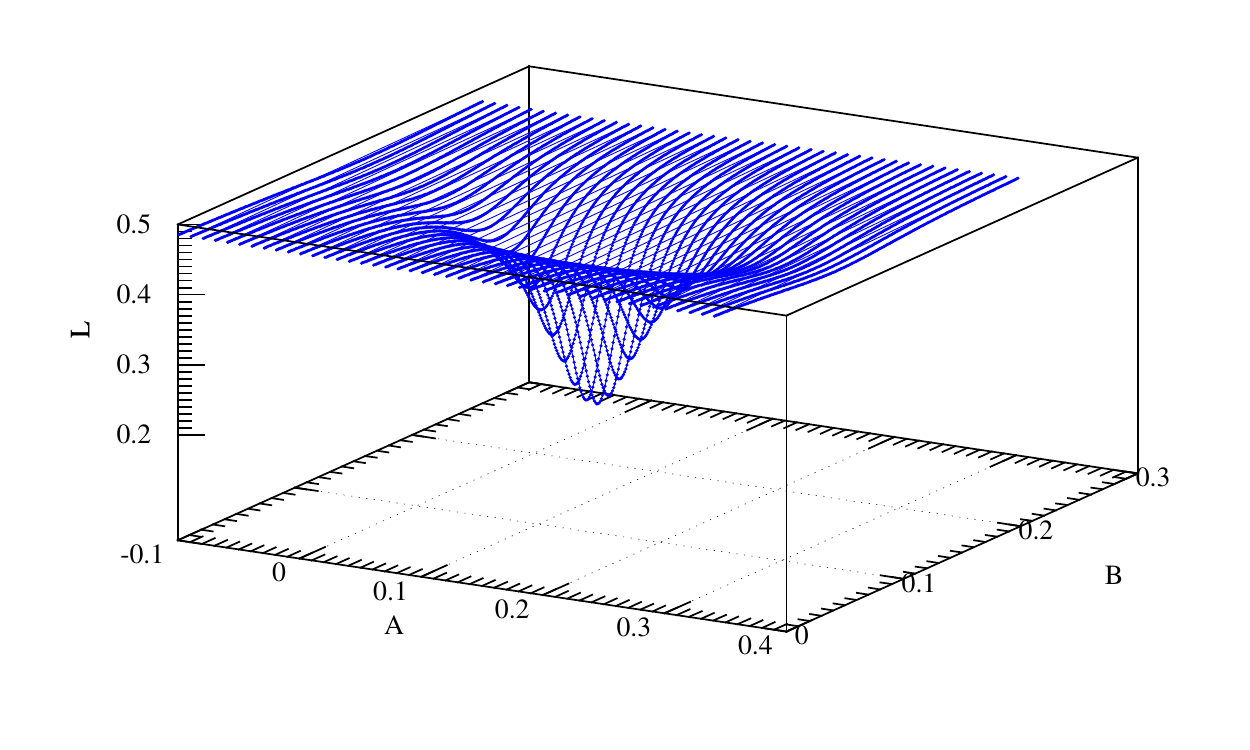}
        \caption{Numerical determination of the minimum of $L$ in the $AB$ plane.}\label{abl}
    \end{minipage}\hspace{5mm}
    \begin{minipage}{0.45\textwidth}
        \centering
        \includegraphics[width=1.1\linewidth,clip]{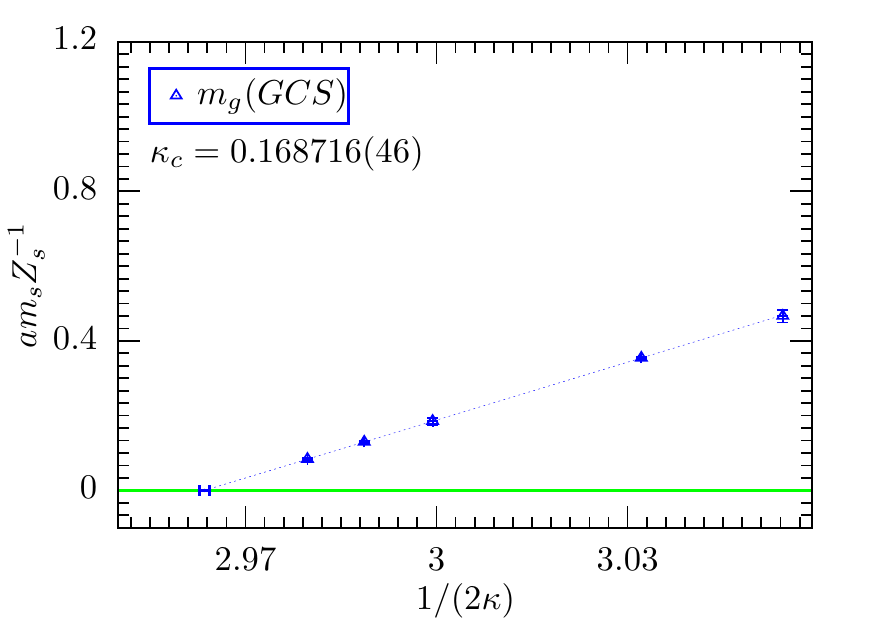}
        \caption{Extrapolation to the chiral limit using gluino masses from the
	Generalized least squares method.}\label{KcGCS}
    \end{minipage}
\end{figure}
Figure \ref{abl} represents the global minimum of $L$ for a numerical range of $A$ 
and $B$. This minimum corresponds to the sought-after values of $A$ and $B$. We re-sample 
the data and use the jackknife method to get the statistical errors. The value of $B$, i.\,e.\ the gluino 
mass ($am_SZ_S^{-1}$), is more or less the same as we already have obtained from the Global 
method, but this time we have a precise and reliable estimate on errors. The method is repeated for the whole range 
of $\kappa$, and we again obtain the critical point $\kappa_c$ (figure \ref{KcGCS}).
 

\subsection{Adjoint pion mass}
 
The mass of the adjoint pion ($m_{\api}$) is an another way to define the critical point, which can be measured 
from the exponential decay of the connected part of the $\aetap$ correlator. The $\api$ is an unphysical particle 
in supersymmetric Yang-Mills theory. However, by arguments based on the OZI-approximation 
\cite{Veneziano:1982ah}, and in the framework of partially quenched chiral perturbation theory~\cite{Munster:2014cja},
the squared mass $m_{\api}^2$ is expected to vanish linearly with the gluino mass 
close to the chiral limit. The 
value of $\kappa_c$ based on $m_{\api}$ can easily be obtained in the simulations. The mass
$m_{\api}$ is usually used to tune $\kappa_c$, given its precision and its cheap computational cost. \par

\begin{figure}[h]
	\centering
	\includegraphics[width=0.8\textwidth,clip]{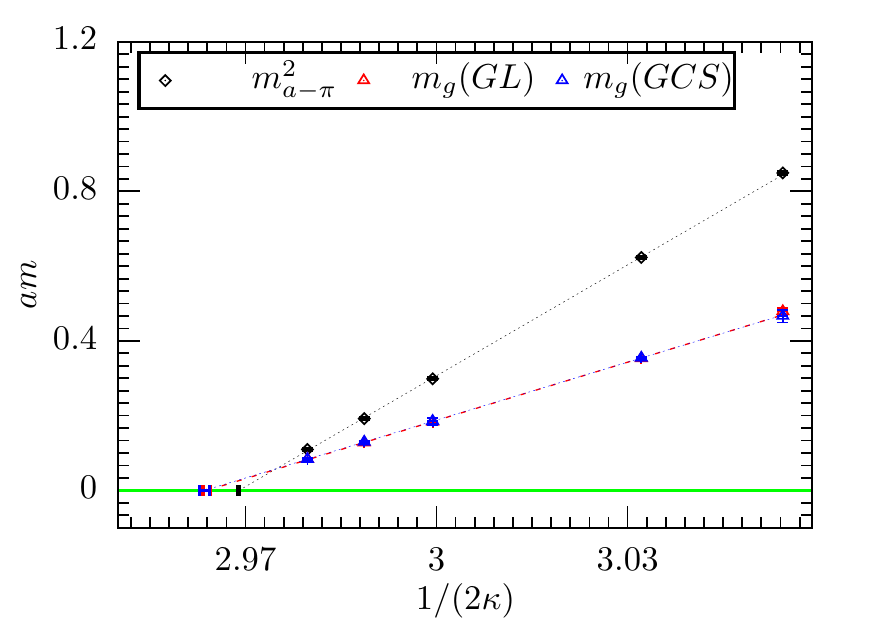}
	\caption{Extrapolation towards the chiral limit ($\kappa_c$) by $m_{\api}$ and by 
	SUSY Ward identities, using both the Global and Generalized least squares methods.}\label{Kcall}
\end{figure}

In figure~\ref{Kcall} we show a comparison 
between $\kappa_c$ obtained independently from the SUSY Ward identities and from $m_{\api}$. The 
values of $\kappa_c$ obtained from the Ward identities and from $m_{\api}$ are very 
close to each other, but there is a small difference.
The reason for this discrepancy are presumably lattice artifacts, and we expect this 
discrepancy to disappear in the continuum limit.
 

\section{Summary and conclusions}
The quantities $ A = Z_TZ^{-1}_S $ and $ B = am_SZ^{-1}_S $(renormalized gluino mass) have been computed 
numerically on the lattice in $\mathcal{N}=1$ SU(3) SYM from on-shell supersymmetric Ward identities. 
The non-perturbative determination of $am_SZ^{-1}_S$ can be used to obtain the chiral point ($\kappa_c$) 
from the Global and the Generalized least squares methods, which are completely in agreement. In comparison 
to the values of gluino masses for gauge group SU(2) presented in~\cite{Bergner2016}, we get more precise results corresponding to each 
value of $\kappa$. The value of the chiral point from Ward identities is compared to the value obtained 
independently from the adjoint pion mass. Both values are compatible up to lattice artifacts. The results 
are consistent with the restoration of supersymmetry in the continuum limit.
 

\section*{Acknowledgements}
The authors gratefully acknowledge the Gauss Centre for Supercomputing (GCS) for providing computing time for a GCS Large-Scale Project on the GCS share of the supercomputer JUQUEEN at J\"ulich Supercomputing Centre (JSC). GCS is the alliance of the three national supercomputing centres HLRS (Universit\"at Stuttgart), JSC (Forschungszentrum J\"ulich), and LRZ (Bayerische Akademie der Wissenschaften), funded by the German Federal Ministry of Education and Research (BMBF) and the German State Ministries for Research of Baden-W\"urttemberg (MWK), Bayern (StMWFK) and Nordrhein-Westfalen (MIWF). Further computing time has been provided on the supercomputers JURECA at JSC, SuperMUC at LRZ, and on the compute cluster PALMA of the University of M\"unster. 

\clearpage

\end{document}